\documentclass[a4paper]{jpconf}
\usepackage{graphicx}
\begin{document}
\title{Thermoelectric transport of perfectly conducting channels
in two- and three-dimensional topological insulators}

\author{Shuichi Murakami$^{1,2}$, Ryuji~Takahashi$^1$,
O.~A.~Tretiakov$^3$, Ar.~Abanov$^3$, Jairo~Sinova$^3$}

\address{$^1$ Department of Physics, Tokyo Institute of Technology, 
2-12-1 Ookayama, Meguro-ku, Tokyo 152-8551, Japan}
\address{$^2$  
PRESTO, Japan Science and Technology Agency (JST), Kawaguchi, Saitama 332-0012, Japan}
\address{$^3$ 
Department of Physics, Texas A\&M University, College Station, Texas 77843, USA}
\ead{murakami@stat.phys.titech.ac.jp}

\begin{abstract}
Topological insulators have gapless edge/surface states with novel
transport properties.  Among these, there are two classes of perfectly
conducting channels which are free from backscattering: the edge
states of two-dimensional topological insulators and the
one-dimensional states localized on dislocations of certain three-dimensional
topological insulators. 
We show how these novel states affect
thermoelectric properties of the systems and discuss possibilities to
improve the thermoelectric figure of merit using these materials with
perfectly conducting channels.
\end{abstract}

\section{Introduction}

Topological insulators (TI)
\cite{Kane05a,Kane05b,Bernevig06a,Wu06,Xu06a} have attracted a lot of
attention in the recent years. In these materials the excitations in
the bulk are gapped just like in the normal insulators, but due to
specific topological properties of the band structure these insulators
possess gapless modes at the edges (on the surface). These modes are
topologically protected and are stable against any perturbations that
do not break the time-reversal symmetry.  Various novel properties of
TIs have been theoretically proposed, and some of them have been
experimentally observed.  The advent of this class of materials has
brought us renewed interest in nonmagnetic insulators which have
rarely been considered as an interesting subject for studying
electronic properties.
The transport properties of the edge/surface excitations in
topological insulators are very distinct from those of the low-energy
excitations of conventional metals.  In two-dimensional (2D)
topological insulators, the edge states are free from elastic
scattering by nonmagnetic disorder. In three-dimensional (3D)
topological insulators \cite{Fu07b,Moore07}, elastic scattering for
the surface states is at least partly suppressed.  It is also proposed
that in certain classes of 3D topological insulators, the lattice
dislocations have gapless states, which are robust against elastic
backscattering by nonmagnetic disorder \cite{Ran}.

Thus, we have two cases of topologically protected one-dimensional
(1D) gapless states in topological insulators; (i) the edge states in
2D topological insulators, and (ii) the states bound to dislocations
in certain 3D topological insulators. These states have similar
physical properties, that they are perfectly conducting, and they
carry spin current. Hence these 1D gapless states are expected to have
large contributions to transport.  Having in mind that some of the
topological insulators are good thermoelectric materials, we studied
the thermoelectric properties of these 1D states, i.e. for edge states
\cite{Takahashi} and for states bound to dislocations
\cite{Tretiakov}. In the present paper, we review the results of these
articles and discuss them in detail. The paper is organized as
follows.  In Section 2 we study thermoelectric transport by the edge
states in 2D topological insulators, following \cite{Takahashi}.
Section 3 is devoted for thermoelectric transport by the states bound
to dislocations in 3D topological insulators \cite{Tretiakov}. The
results are summarized in Section 4.

Thermoelectric properties of the system 
are characterized by the thermoelectric figure of merit $ZT$ defined by
\begin{eqnarray}
ZT&=&\frac{\sigma S^2 T}{\kappa},
\end{eqnarray}
where $\sigma$ is the charge conductivity, $S$ is the Seebeck
coefficient, $\kappa$ is the thermal conductivity, and $T$ is the
temperature.  The larger $ZT$ means better efficiency for
thermoelectric conversion. There are various approaches toward
improving the $ZT$. In many materials, the main obstacle for better
$ZT$ is a large phonon heat transport. One approach to impove $ZT$ is
called phonon-glass electron-crystal (PGEC) \cite{Slack}, i.e. a good
conductor for electrons but a poor conductor for phonons.  The phonon
conduction is suppressed by varying the lattice structure, and thus
making the system glassy for phonons. Nevertheless the electron
conduction should remain good for a good thermoelectric transport.
Another approach for improving $ZT$ is low-dimensionality
\cite{Hicks93a}.  This idea is to control the electronic band
structure, so that the density of states has a peaked behaviour
\cite{Mahan96}.  Despite these attempts the maximum $ZT$ remains of
order of unity, awaiting for a breakthrough.

\section{Thermoelectric transport by edge states of 2D topological insulators}

Because the 2D topological insulators have gapless edge states which
are free from elastic scattering, they might have a possibility for
novel and enhanced transport properties.  For charge transport it has
been measured experimentally in HgTe quantum well.  In multiterminal
measurement, the edge channels are shown to be perfectly conducting
\cite{Konig07,Roth}.  While the edge states form perfectly conducting
channels, the number of edge states is typically much smaller than
that of the bulk, and the overall transport is mainly contributed by
the bulk.  In charge transport, we can neglect the bulk transport at
lower temperatures, if the chemical potential is within the bulk gap.
Nevertheless, because we are interested in the heat transport, we
should consider finite temperatures, and bulk carriers cannot be
neglected.  Hence, in order to make the edge states more prominent
compared with the bulk, we consider the system in narrow ribbon
geometry.

The charge and heat current in a solid is described by the following
linear response,
\begin{eqnarray}
\left(\begin{array}{c}
 j/q \\ w \end{array}\right)
=
\left(\begin{array}{cc} L_{0} & L_{1} \\ L_{1} & L_{2} \end{array}\right)
\left(\begin{array}{c} -\nabla\mu \\ -T^{-1}\nabla T \end{array}\right),
 \label{eq:yusougyouretsu}
\end{eqnarray}
where $j$ is the current density, $q$ is the electron charge, $w$ is the heat current density, 
and $\mu$ is the chemical potential.
$L_i$ are the transport
coefficients, using which the various physical quantities are expressed as
\begin{equation}
\sigma=q^2 L_0,\ \ 
S =\frac{L_1}{qTL_0},\ \ 
\pi =\frac{L_1}{qL_0},\ \ 
\kappa= \kappa_{\rm{ph}}+\kappa_{\rm{e}},\ \ \kappa_e = \frac{1}{T}\frac{L_0 L_2 - L^2_1}{ L_0}
\end{equation}
where $\kappa_{\rm{ph}}$ and $\kappa_{\rm{e}}$ are the thermal
conductivity by the phonons and that by the electrons, respectively.

\subsection{Thermoelectric transport by edge states only}
\begin{figure}[htbp]
 \begin{center}
  \includegraphics[width=100mm]{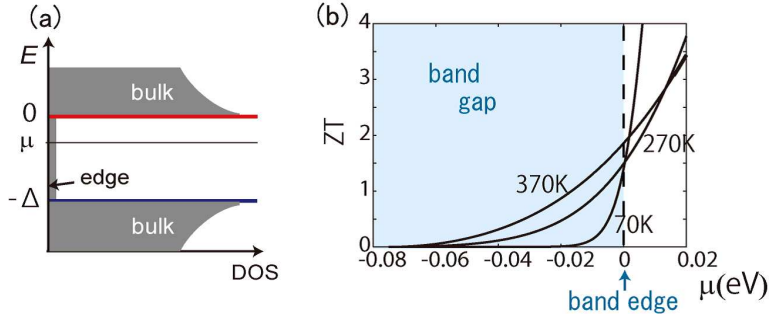}
 \end{center}
 \caption{(a) Schematic density of states (DOS). (b) Figure of merit
   $ZT$ as a function of chemical potential $\mu$, calculated from
   edge transport only. Bulk carriers and phonons are
   neglected. $\Delta$ is taken to be $\Delta=0.15$eV.}
 \label{fig:edge}
\end{figure}

There are three carriers which contribute to $ZT$: the edge electrons,
bulk electrons, and phonons.  We first consider an ideal situation of
only edge states, while transport by the bulk electrons and phonons is
neglected.  The density of states is schematically described as in
Fig.~\ref{fig:edge}(a).  We set the origin of the energy as the
conduction band edge.  We apply the Landauer-B\"{u}ttiker formula
\begin{eqnarray}
L_{\nu}=\frac{1}{h}\int\sum_{i}\mathcal{T}_i(E)(E-\mu)^{\nu}\left(-\frac{\partial f}{\partial E}\right)\mathrm{d}E,
\end{eqnarray}
where $i$ is the channel index, $\mathcal{T}_{i}(E)$ is the
transmission probability for the $i$-th channel at energy $E$, and
$f(E)=1/(e^{(E-\mu)/k_{\rm B}T}+1)$ is the Fermi distribution function
\cite{Sivan}.  In the present case, the transmission probability is
equal to unity for the states with energy $E$ at $-\Delta<E<0$, and
zero otherwise, with $\Delta$ being the size of the bulk gap. We then
obtain
\begin{eqnarray}
L^{{\rm e}}_\nu &=&\frac{2\ell}{sh}\int^{0}_{-\Delta}(E-\mu)^{\nu}\left(-\frac{\partial f}{\partial E}\right)\mathrm{d}E 
=c_eF_{\rm e}(\nu,\mu)(k_B T)^\nu,\\
&\ &\ \ \ \ c_{\rm e}=\frac{2\ell}{sh},\ \ F_{\rm e}(\nu,\mu)=\int^{-\bar{\mu}}_{-\bar{\Delta}- \bar{\mu}} x^{\nu} \frac{\mathrm{e}^x}{(\mathrm{e}^x+1)^2}  \mathrm{d}x,
\label{eq:edge}
\end{eqnarray}
where $s$ is the cross section of the ribbon, $\ell$ is an effective
length of the ribbon, $\bar \Delta= \Delta/k_B T$, and $\bar \mu=
\mu/k_B T$.  The superscript ``e" denotes the edge transport.  The
effective length $\ell$ is taken as the inelastic scattering length of
the edge states $\ell_{\mathrm{inel}}$.  If we neglect the phonon
transport, the figure of merit $ZT$ is
\begin{eqnarray}
ZT &=& \frac{L_1^2}{L_0 L_2 - L_1^2}
\end{eqnarray}
The resulting $ZT$ is shown in Fig.~\ref{fig:edge}(b).  We can see
that as the chemical potential becomes larger than zero and goes into
the bulk band, the $ZT$ increases rapidly.  Compared with the $ZT$
ever achieved in experiments which is around $ZT\sim 2$, the resulting
$ZT$ is extremely large.  It is because $ZT$ reflects the average
energy per carrier in the unit of $k_{\rm B}T$; if the chemical
potential is in the bulk band, the edge carriers have larger energy in
average, giving large $ZT$.

In this regime of $\mu$, the edge carriers are thermally excited holes. 
Therefore the carrier numbers are exponentially small, so are the transport coefficients $L^{\rm e}_{i}$. 
Hence, if we include bulk carriers or phonons, they will easily become dominant over the edge contributions,
suppressing the large $ZT$.

\subsection{Competition between edge and bulk transport}
\begin{figure}[h]
 \begin{center}
  \includegraphics[width=120mm]{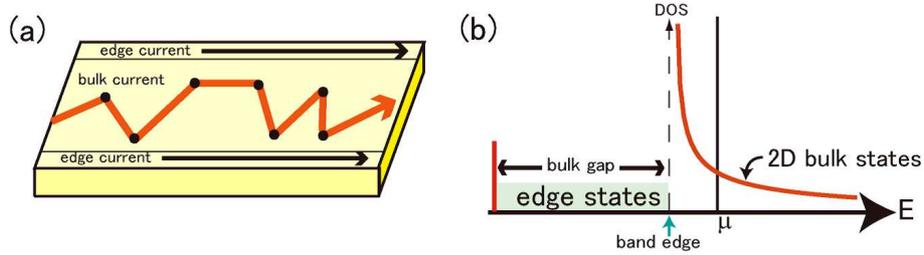}
 \end{center}
 \caption{(a) Schematic picture for the two-dimensional ribbon. (b)
   Density of states including edge states and bulk states at the
   first subband.}
 \label{fig:parasanran}
\end{figure}

In addition to edge transport, we consider the bulk transport. The
schematic figure is shown in Fig.~\ref{fig:parasanran}.  For
simplicity, we set the bulk gap to be sufficiently larger than the
temperature, and the chemical potential $\mu$ to be near the
conduction band edge. Under these assumptions the bulk valence band is
neglected.  We also note that we are considering very narrow ribbons,
for which the transverse motion is quantized due to the
confinement. For ribbons as narrow as $W=10$nm wide or so, we can
assume that only the first subband is involved.

\begin{figure}[htbp]
 \begin{center}
  \includegraphics[width=110mm]{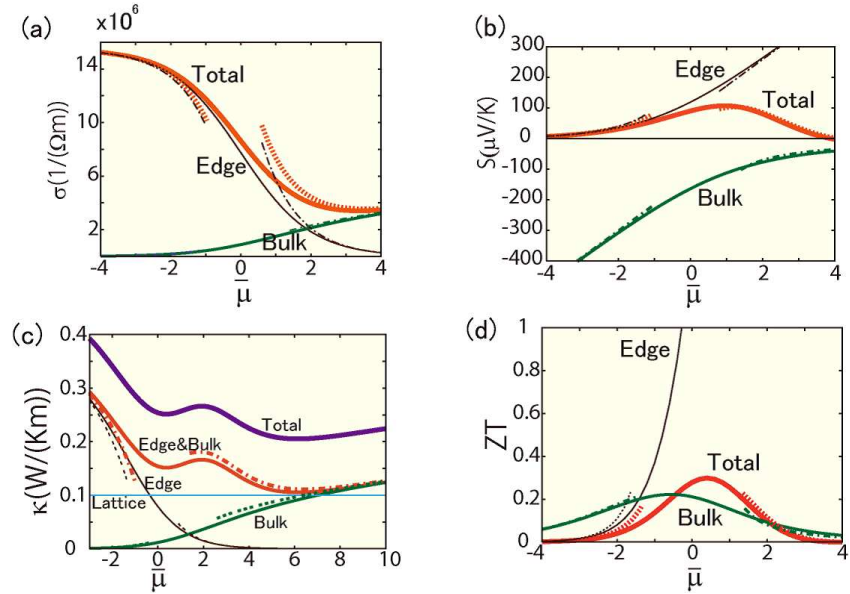}
 \end{center}
 \caption{(a) Charge conductivity $\sigma$, (b) Seebeck coefficient
   $S$, (c) thermal conductivity $\kappa$, (d) thermoelectric figure
   of merit $ZT$. The horizontal axis represents $\bar{\mu}=\mu
   /k_{B}T$, and $\bar{\mu}=0$ corresponds to the band edge.}
 \label{fig:sskapZT}
\end{figure}

Bulk transport is diffusive and calculated by the Boltzmann transport theory. For simplicity we
put the lifetime $\tau$ to be constant.
\begin{eqnarray}
&&L^{{\rm b}}_\nu =\frac{1}{s}\int_{-\infty}^{\infty}\frac{\mathrm{d}k}{2\pi}~c\tau v^{2}(\epsilon(k)-\mu)^{\nu}\left(-\frac{\partial f(\epsilon(k))}{\partial \epsilon}\right)
=\frac{4\sqrt{2m}\mu^{*}c}{esh}\int_{0}^{\infty}\mathrm{d}\epsilon~\sqrt{\epsilon}(\epsilon-\mu)^{\nu}
\left(-\frac{\partial f}{\partial \epsilon}\right)
\nonumber \\
&&=c_bF_{\rm b}(\nu,\mu)(k_B T)^\nu,
\label{eq:Lb}\\
&&c_{\rm b}=\frac{4\sqrt{2mk_B T}\mu^{*}c}{esh}, \ \ 
F_{\rm b}(\nu,\mu) =\int^{\infty}_{-\bar{\mu}}\sqrt{x+\bar{\mu}}\ x^\nu\frac{ 
 \mathrm{e}^{x}}{(\mathrm{e}^{x }+1)^2}  \mathrm{d}x,
\label{eq:bulk}
\end{eqnarray}
where $\epsilon(k)=k^2/2m$, $c$ is the number of carrier pockets,
$\mu^{*}$ is the mobility, $m$ is the effective mass, and the
superscript ``b" denotes the bulk transport.  The prefactors $c_{\rm
  e}$ and $c_{\rm b}$ characterize the size of the transport for the
edge and the bulk, respectively, and the total transport coefficient
is given by their sum:
\begin{eqnarray}
L_\nu=L^{{\rm e}}_\nu+L^{{\rm b}}_\nu.
\end{eqnarray}
The results for the transport coefficients at $T=1.8$K are shown in
Fig.~\ref{fig:sskapZT} (a)-(d). We took Bi$_2$Te$_3$ as an example
\cite{Chen10}, and adopt the parameters as $c=6$,
$\ell_{\mathrm{inel}}=1\mu$m, $m^*=0.02m_e$.  The cross section of the
ribbon is set to be $s=10{\rm nm}\times$ 0.5 nm.  The mobility is
$\mu^{*}=2000cm^2$V$^{-1}$s$^{-1}$, taken from the bulk mobility at
80K \cite{Champness}.  For the phonon thermal conductivity we assumed
$\kappa_L=0.1 $Wm$^{-1}$K$^{-1}$ estimated from the bulk values
\cite{Macdonald,Balandin}.

In Fig.~\ref{fig:sskapZT}, we observe interesting interplay between the edge and bulk transport. 
It is reasonable that 
in the charge conductivity (Fig.~\ref{fig:sskapZT}(a)) and the thermal conductivity (Fig.~\ref{fig:sskapZT}(c)),
when the chemical potential is in the bulk gap ($\bar{\mu}\ll 0$), the transport is dominated by 
edge carriers. On the other hand, when the chemical potential is in the bulk band ($\bar{\mu}\gg 0$), the transport is dominated by 
bulk carriers.
The behavior of the Seebeck coefficient $S$ (Fig.~\ref{fig:sskapZT}(b)) is interesting. First, the edge and bulk contributions
have opposite signs. It is because the sign of the Seebeck coefficient is the same as the sign of the carriers involved.
When the chemical potential is near the bulk band edge, the carriers in the bulk conduction band are electrons, whereas the 
carriers in the edge states are holes. Thus they have opposite contributions for the Seebeck coefficients. 
It is also noted that the Seebeck coefficient from the edge states is an increasing function of $\mu$, while that from 
the bulk states is a decreasing function of $\mu$. They are both asymptotically linear functions.

As a whole, the figure of merit $ZT$ (Fig.~\ref{fig:sskapZT}(d)) has a
maximum around $\mu=0$, the bulk band edge.  We note that although the
$ZT$ from the edge is very large when $\mu$ is large, it is reduced
drastically by the bulk carriers and phonons. It is because in this
regime of $\mu$, the carrier concentration of edge states is
exponentially small, and it can be easily overcome by the bulk
carriers and the phonons.

We can evaluate the transport coefficients when $\bar{\mu}$ is far from the band edge, i.e. $|\bar{\mu}|\gg 1$. 
Here we also neglect the valence band and focus only on the conduction band and edge states, 
by setting $\bar{\Delta}=\Delta/(k_{\rm B}T)\rightarrow \infty$.
For the edge states, when $\bar{\mu}\gg 1$ (chemical potential is deep inside the bulk conduction band), 
the asymptotic form of $F_{\rm e}(\nu,\mu)$ is
\begin{equation}
F_{\rm e}(0,\mu)\sim\mathrm{e}^{-\bar{\mu}}, \ \  
F_{\rm e}(1,\mu)\sim-(1+\bar{\mu})\mathrm{e}^{-\bar{\mu}},\ \ 
F_{\rm e}(2,\mu)\sim(2+2\bar{\mu}+\bar{\mu}^2)\mathrm{e}^{-\bar{\mu}},
\end{equation}
and when $\bar{\mu}\ll -1$ (chemical potential is deep inside the bulk gap), 
\begin{equation}
F_{\rm e}(0,\mu)\sim1-\mathrm{e}^{\bar{\mu}},\ \ 
F_{\rm e}(1,\mu)\sim(\bar{\mu}-1)\mathrm{e}^{\bar{\mu}},\ \ 
F_{\rm e}(2,\mu)\sim\frac{\pi^2}{3}+(-\bar{\mu}^2+2\bar{\mu}-2)\mathrm{e}^{\bar{\mu}}.
\end{equation}
On the other hand, 
for the bulk states, when $\bar{\mu}\gg 1$ (chemical potential is deep inside the bulk conduction band), 
the asymptotic form of $F_{\rm b}(\nu,\mu)$ is
\begin{equation}
F_{\rm b}(0,\mu)\sim\sqrt{\bar{\mu}},\ \ 
F_{\rm b}(1,\mu)\sim\frac{\pi^2}{6\sqrt{\bar{\mu}}},\ \ 
F_{\rm b}(2,\mu)\sim\frac{\pi^2}{3}\sqrt{\bar{\mu}}
\end{equation}
and when $\bar{\mu}\ll -1$ (chemical potential is deep inside the bulk gap), 
\begin{equation}
F_{\rm b}(0,\mu)\sim\frac{\sqrt{\pi}}{2}\mathrm{e}^{\bar{\mu}},\ \ 
F_{\rm b}(1,\mu)\sim\frac{\sqrt{\pi}}{2}\mathrm{e}^{\bar{\mu}}(\frac{3}{2}-\bar{\mu}),\ \ 
F_{\rm b}(2,\mu)\sim\frac{\sqrt{\pi}}{2}\mathrm{e}^{\bar{\mu}}(\frac{15}{4}-3\bar{\mu}+\bar{\mu}^2).
\end{equation}
From these asymptotic forms, we evaluate the asymptotics for the transport properties, shown 
in Fig.~\ref{fig:sskapZT} as broken lines.
In particular, the $ZT$ from the edge state only has the asymptotic form
\begin{eqnarray}
&&ZT_{\rm e}\sim (1+\bar{\mu})^2: \bar{\mu}\gg 1,\\
&&ZT_{\rm e}\sim \frac{3}{\pi^2}(1-\bar{\mu})^2e^{\bar{\mu}}: \bar{\mu}\ll -1,
\end{eqnarray}
and
the $ZT$ from the bulk state only has the asymptotic form
\begin{eqnarray}
&&ZT_{\rm b}\sim \frac{\pi^2}{12\bar{\mu}^2}: \bar{\mu}\gg 1, \\ 
&&ZT_{\rm b}\sim \frac{2}{3}\left(\frac{3}{2}-\bar{\mu}\right)^2: \bar{\mu}\ll -1.
\end{eqnarray}

\subsection{Universal behavior described by dimensionless parameters}
\begin{figure}[hbp]
 \begin{center}
  \includegraphics[width=90mm]{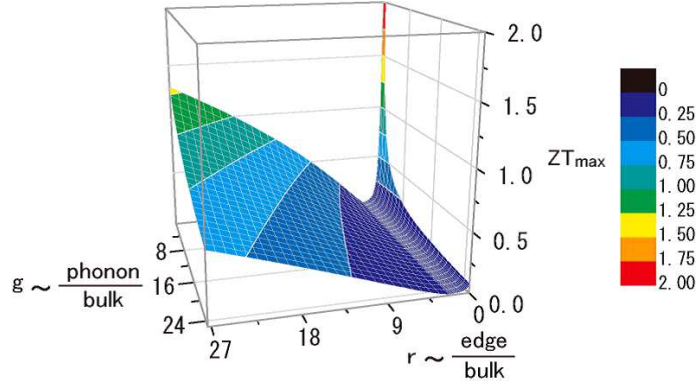}
 \end{center}
 \caption{$ZT_{\rm max}$ as a function of $r$ and $g$. $\bar{\mu}$ is set to maximize $ZT$.}
 \label{fig:ZT3D}
\end{figure}

We have calculated various thermoelectric properties using specific values of parameters.
There is a wide range of choices for various parameters, such as sample geometry, 
material choice, temperature, and so forth. 
It is therefore important to develop a theory which is applicable for 
various choices of parameters. This can be achieved by introducing dimensionless 
parameters $r$ and $g$ defined by
\begin{equation}
r=c_{\rm e}/c_{\rm b}
=
\frac{e\ell}{2\sqrt{2mk_B T}\mu^{*}c}, \ \ \ 
g=\left(\frac{\kappa_L}{k_B^2 T}\right)/c_{\rm b}
=\frac{\kappa_L esh}{4\sqrt{2mk_B^{5}T^{3}}\mu^{*}c}.
\end{equation}
The parameter $r$ characterizes the ratio of edge transport to bulk
transport, and $g$ characterizes the phonon transport to bulk
electronic transport.  The figure of merit $ZT$ is then expressed as a
function of $r$, $g$ and $\bar{\mu}$.  We then maximize $ZT$ as a
function of $\bar{\mu}$. In Fig.~\ref{fig:ZT3D} we show the maximum
value of $ZT$ for each value of $r$ and $g$.
maximum.  One can see from Fig.~\ref{fig:ZT3D} that $ZT$ has a minimum
at $r\sim 2.3$.  As we increase $r$, the edge states become more and
more dominant, and bulk-to-edge crossover occurs. Because the edge and
bulk tend to cancel each other, this crossover is accompanied by a
minimum of $ZT$.  In general the behavior of $r$ and $g$ as functions
of the temperature $T$ is complex.  Nevertheless, the inelastic
scattering length $\ell$ is a rapidly decreasing function of $T$, as
is seen from the case of quantum Hall effect with
$\ell_{\rm{inel}}\sim T^{-1.5} $ \cite{Machida}, and we can estimate
$r\sim T^{-2}$.  Thus we conclude that by lowering temperature, edge
states becomes dominant, and there occurs a bulk-to-edge
crossover. The $ZT$ thereby has a minimum and then increases again as
a function of temperature. For 10nm-wide ribbon, the crossover is
expected to be around 5K-10K.

\subsection{Discussion}
In the above estimate, the value of $ZT$ is sensitive to the
edge-state inelastic scattering length $\ell$. If $\ell$ becomes
longer, it is more favorable for edge-state transport since it is free
from scattering up to a longer distance.  It is difficult to
qualitatively estimate $\ell$ in topological insulators because it is
dominated by electron-electron scattering. We can draw analogy with
quantum Hall systems, where $\ell$ is estimated to be 1$\mu$m at 1K
\cite{Machida}. On the other hand, in HgTe quantum well, known as a 2D
topological insulator, perfectly conducting channels for the system
size 1$\mu$m are observed at 1.8K, meaning that $\ell>1\mu$m
\cite{Konig07}. From these data we used $\ell =1{\mu}{\rm m}$ in our
calculation.

Because $r$ is inversely proportional to the ribbon width, narrower
ribbons are more favorable for edge-dominated transport, in agreement with
our expectations. On the other hand, there is a minimum width for which
the ribbon supports the perfectly conducting edge channels. It is
determined by a penetration depth $\lambda$ of edge states.  If the
ribbon width is narrower than the penetration depth of edge states,
there occurs a backscattering between the edge states on the opposite
sides of the ribbon, and the edge state transport is suppressed.  In
HgTe quantum well the penetration depth is 6nm-50nm
\cite{Zhou,Wada10}. On the other hand, in some systems such as
bismuth(111) 1-bilayer films, $\lambda$ is predicted to be as short as
a lattice constant \cite{Murakami06,Wada10}, and the ribbon can be
made very narrow with keeping the perfectly conducting edge channels.

\section{Thermoelectric transport by dislocations in 3D topological insulators}

Compared with 2D topological insulators where the edge states are free
from scattering, the surface states of 3D topological insulators
undergo some scattering.  In a case of Dirac cone as seen in
Bi$_2$Se$_3$, the perfect backscattering by 180 degrees is prohibited,
but other scattering processes can happen. Therefore, the surface
states do not show perfect conduction.

\begin{figure}[hbp]
 \begin{center}
  \includegraphics[width=110mm]{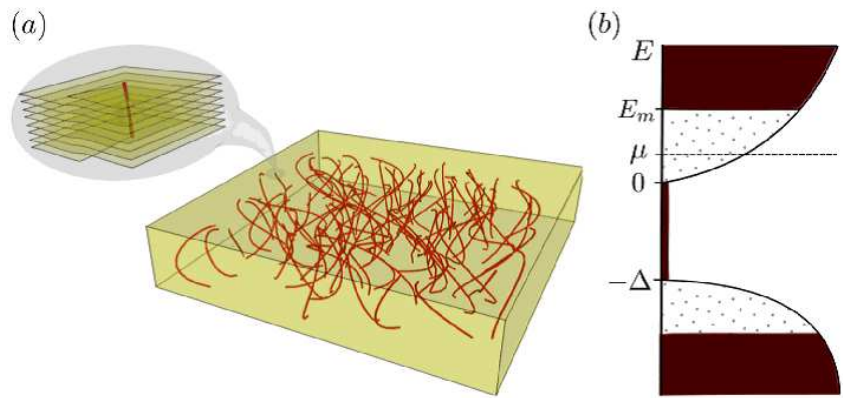}
 \end{center}
 \caption{(a) Dislocations in 3D topological insulator. The inset
   shows one such dislocation of the screw type. (b) Schematic
   picture of the band structure for the calculation.}
 \label{fig:disloc}
\end{figure}

Nevertheless, it is theoretically predicted that in some of the 3D
topological insulators there occurs a Kramers pair of gapless states
on a lattice dislocation \cite{Ran}. This state carries a pure spin
current and is topologically protected.  Thus these dislocations form
perfectly conducting channels, as is similar to the edge states in 2D
topological insulators.  It is then expected that if we increase the
number of dislocations, as shown in Fig.~\ref{fig:disloc}(a), the
transport by these 1D channels becomes dominant over the bulk
transport and phonon transport.  Increasing the number of dislocations
in the crystal is also helpful for other reasons: the mean free path
of phonons is reduced by disorder, and the bulk carriers localize
below the mobility edge formed at the bottom of the conduction band.
Thus more dislocations enhance the 1D channel transport while at the
same time suppress the bulk electron transport and phonon transport.

\begin{figure}[hbp]
\begin{center}
  \includegraphics[width=130mm]{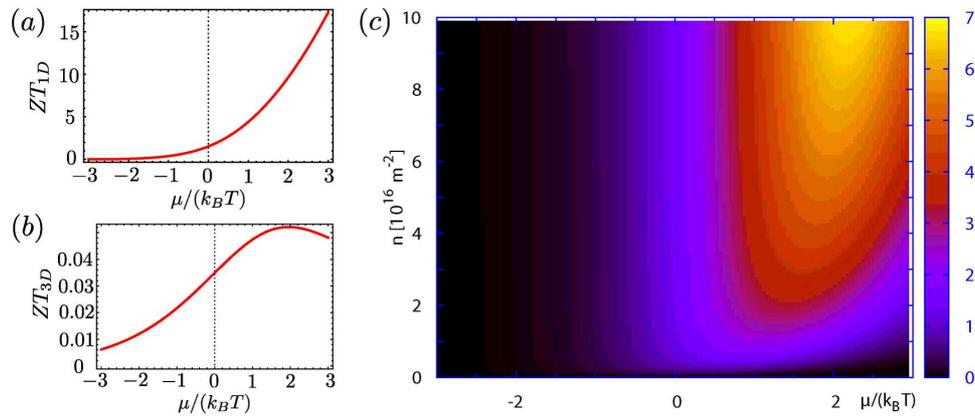}
 \end{center}
 \caption{(a) Figure of merit $ZT_{1D}$ from 1D channels bound to
   dislocations as a function of chemical potential $\mu$. (b) Figure
   of merit $ZT_{3D}$ for the bulk of the 3D topological
   insulator. The 1D channels are neglected here. (c) Contour plot of
   the total $ZT$ as a function of the dislocations density $n$ and
   the chemical potential $\mu$.}
 \label{fig:ZT}
\end{figure}

The framework of the $ZT$ calculation is similar to the 2D ribbon, and
we do not provide it here. The details can be found
in~\cite{Tretiakov}.  All the calculations are performed for room
temperature, $T=$300 K.  We assume that the phonon thermal
conductivity is reduced due to disorder from the bulk value of
$\kappa_{ph}=1{\rm Wm}^{-1}{\rm K}^{-1}$ to $\kappa_{ph}=0.01{\rm
  Wm}^{-1}{\rm K}^{-1}$ for the average distances between dislocations
$\sim 3$ nm. The gap between the valence and conduction bands is taken
to be $\Delta=0.15$ eV. We also set the mobility edge to be $E_m=0.05$
eV, measured from the bottom of the conduction band, see
Fig.~\ref{fig:disloc}(b).  The resulting $ZT$ for the 1D channel and
the 3D bulk carriers are given in Fig.~\ref{fig:ZT}(a) and (b),
respectively. As a function of the concentration of dislocations the
figure of merit $ZT$ is calculated in Fig.~\ref{fig:ZT}(c). The
highest value $ZT$ reaches when the chemical potential $\mu$, which
can be changed by an external gate, is near the mobility edge.
Remarkably the maximum value of $ZT$ is over 6 at room temperature.

It is important to mention here that, if the impurities are
non-magnetic, the transition to a bulk Anderson insulator should not
destroy the 1D gapless states, because the time reversal invariance is
not broken. Note also that by increasing the density of dislocations
and/or non-magnetic impurities one can reach even greater values of
$ZT$. Nevertheless, these densities are less likely to be achieved and
can as well result in the tunneling of electrons between the channels
of neighboring dislocations. This tunneling could lead to the gap
opening and localization of the protected 1D states. This is the
reason for our choice of an upper limit of the dislocation density of
$n\sim 10^{17}$m$^{-2}$ which corresponds to an average dislocation
spacing of around 3 nm.

We note that not every topological insulator has dislocations carrying
1D gapless protected modes.  The condition for the 1D gapless
protected modes is expressed as ${\bf B}\cdot{\bf M}_{\nu}=\pi\ ({\rm
  mod} 2\pi)$, where the vector ${\bf M}_{\nu}$ (so called
time-reversal-invariant momentum) is a part of the $Z_2$ topological
number for the topological insulator considered and ${\bf B}$ is the
Burgers vector of the dislocation \cite{Ran}.  Therefore, only the
materials which $Z_2$ topological numbers satisfy the above criterion
can carry 1D gapless modes through the dislocations. In this sense, the
dislocations in Bi$_2$Te$_3$ \cite{Hsieh09a,Hsieh09c,Zhang09a,Chen10}
and Bi$_2$Se$_3$ do not carry 1D gapless modes, but the dislocations
in Bi$_{1-x}$Sb$_x$ \cite{Hsieh08,Nishide10} ($0.07<x<0.22$) possess
1D gapless modes. 

\section{Conclusion}
In the present paper, we consider two types of topologically protected
1D gapless states in 2D and 3D topological insulators, as
current-carrying paths for thermoelectric transport. In both cases,
the 1D modes alone give large thermoelectric figure of merit $ZT$,
whereas the bulk states and phonons suppress the $ZT$.  Thus in order
to have good $ZT$, it is necessary to have high density of 1D modes.
It is accomplished by making narrow ribbons for 2D topological
insulators, and by making dislocations densely distributed in 3D
topological insulators.

\ack
This work is supported partly by Grant-in-Aid for Specially Promoted Research
(21000004), by Grant-in-Aid for Scientific ResearchiC) (22540327), 
 by Grant-in-Aid for Global COE Program "Nanoscience and Quantum 
Physics" from the Ministry of Education, Culture, Sports, Science
and Technology of Japan, 
by NSF under Grant Nos. DMR-0547875 and 0757992, by the
Research Corporation Cottrell Scholar Award, and by the
Welch Foundation (A-1678).

\end{document}